\documentclass[aop,apl,twocolumn,groupedaddress,showkeys,showpacs]{revtex4-1}

\usepackage{graphicx}
\usepackage{ulem}       
\usepackage{color,soul} 
\usepackage{mathtools}  
\usepackage{amssymb}  
\usepackage{epstopdf}

\begin{document}


\title{Size-independent Young's modulus of inverted conical GaAs nanowire resonators}

\author{P. Paulitschke}
\author{N. Seltner}
\author{A. Lebedev}
\author{H. Lorenz}
\author{E.~M. Weig}
\altaffiliation{present address: Universit\"{a}t Konstanz, Universit\"{a}tsstr. 10, 78457 Konstanz, Germany}
\email{eva.weig@uni.konstanz.de}
\affiliation{Center for NanoScience \& Fakult\"{a}t f\"{u}r Physik, Ludwig-Maximilians-Universit\"{a}t, Geschwister-Scholl-Platz 1, 80539 M\"{u}nchen, Germany}

\date{\today}

\begin{abstract}
We explore mechanical properties of top down fabricated, singly clamped inverted conical GaAs nanowires. Combining nanowire lengths of $2-9\,\mu$m with foot diameters of $36-935$\,nm yields fundamental flexural eigenmodes spanning two orders of magnitude from $200$\,kHz to $42$\,MHz.
We extract a size-independent value of Young's modulus of $(45\pm 3)$\,GPa. With foot diameters down to a few tens of nanometers, the investigated nanowires are promising candidates for ultra-flexible and ultra-sensitive nanomechanical devices.
\end{abstract}



\maketitle

The rise of nanotechnologies in basic research as well as the applied sciences goes along with the development of more and more compact and sensitive devices. For example, nanomechanical systems (NEMS) are promising candidates for ultra-responsive mass~\cite{bib:Chaste2012,bib:Hanay2012a}, force~\cite{bib:Mamin2001,bib:Arlett2006,bib:Hallstrom2010} or biosensors~\cite{bib:Burg2007,bib:Waggoner2007,bib:Arlett2011}, as well as accelerometers~\cite{bib:Krause2012a} or oscillators~\cite{bib:Feng2008}. The successful realization of such devices is enabled by a combination of three key factors: integrated architectures, reliable fabrication and high sensitivity. Particularly for integrated sensing devices the vertical arrangement of dense arrays of micro- or nanowires~\cite{bib:Hallstrom2010} is considered beneficial. A compact sensor design with higher functionality and reproducible control over device parameters is facilitated by top down fabrication. Finally, the device's sensitivity is closely related to its spring constant, which is a function of both geometry and material properties such as Young's modulus $E$, as, quite generally, a softer spring allows resolving smaller signals. Thus, detailed knowledge of the mechanical properties is crucial to predict the performance of a device. Typically, Young's moduli are determined by relating the measured resonance frequency $f_0$ of the resonator's fundamental flexural mode to its geometrical dimensions~\cite{bib:Treacy1996,bib:Li2003,bib:Chen2006,bib:Gavan2009a,bib:Lexholm2009,bib:Qin2012}.
A prominent example is the simple singly-clamped cylindrical beam for which Euler Bernoulli beam theory~\cite{bib:Weaver1990} yields the well-known relation
\begin{equation}
\label{cylinder}
2\pi f_0 = 1.758 \sqrt{\frac{E}{\rho}} \frac{r}{h^2}
\end{equation}
for aspect ratios $r/h<0.1$ with mass density $\rho$, beam radius $r$ and length $h$.

Here we present a nanomechanical resonator which is an excellent candidate for a nanomechanical sensing device. Figure~1
shows nanowires etched into an ($100$)-oriented GaAs substrate, combining the benefits of top down fabrication with an ultrasoft mechanical response, allowing for immediate integration into a sensing array. Notably, the nanowires are not cylindrical, but of inverted conical shape. This is apparent from the magnified nanowire in the right part of Fig.~1
featuring a length of $h=6\,\mu$m, a head radius of $R=264$\,nm and a foot radius of only $r=85$\,nm, which corresponds to a taper angle of $\varphi \approx 1.7^\circ$. While the narrow nanowire foots enable high force sensitivities~\cite{bib:Rast2000,bib:Lee2005}, the relatively large nanowire heads can easily be resolved in an optical microscope. As a consequence, inverted conical nanowire arrays are highly promising devices to study e.g. cellular force exertion, as benchmark sensitivities can be obtained without the need to functionalize and thus modify the nanowires using fluorescent markers, which also avoids the frequently encountered bleaching of fluorescent dyes. In addition, the comparatively large heads allow for precise and reproducible nanowire definition
using electron-beam lithography. After the evaporation of a nickel etch mask and a lift-off process, the nanowire is etched using inductive coupled plasma reactive ion etching (ICP RIE) with SiCl$_4$.
Following a thorough conditioning of the etching chamber, nanowires can be fabricated reproducibly with chosen sidewall roughness and taper angle.
For a given taper angle the nanowire length is a function of the etching time and thus relies on the resilience of the etching mask. The presented nanowires have been processed using an rf power of $120$\,W, an ICP power of $90$\,W, $5$\,sccm of SiCl$_4$ flow at a pressure of $0.2$\,mTorr with an etch rate of $470$\,nm$/$min in an Oxford PlasmaLab 100 etcher and feature head radii $R$ between $100$\,nm and $1\,\mu$m. The nanowire length $h$ is varied between $2\,\mu$m and $9\,\mu$m, giving rise to foot radii $r$ from $36$\,nm to $935$\,nm.
\begin{figure}[htb]
\begin{center}
\includegraphics[width=8.25cm]{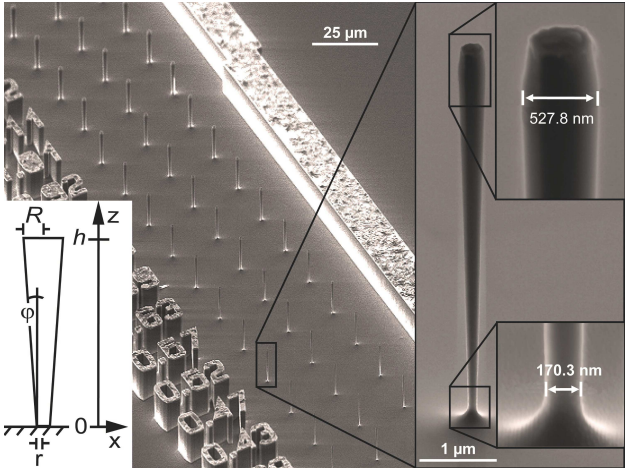}
\caption{Left: Scanning electron micrograph showing a tilted view of a sample hosting rows of flexible nanowires of varying diameter. Lithographed numbers refer to head diameter $2R$ of the respective row of three identical nanowires in micrometers. All nanowires on the displayed chip have the same length $h=6\,\mu$m. Right: Enlarged view of a single nanowire illustrating its inverted conical shape, as well as further close-ups determining the nanowire head and foot diameters $2R$ and $2r$, respectively. The left inset depicts a schematic of the inverted conical nanowire along with the employed nomenclature.}
\end{center}
\label{pillar}
\end{figure}
As a last step, all nanowires are imaged using a high resolution scanning electron microscope (SEM) to determine $R$, $r$ and $h$ with an accuracy of about $5$\,nm resulting from finite pixel size and edge effects in SEM images (see right part of Fig.~1).
Subsequently, the chips are mounted in a vacuum chamber to characterize the nanowires' eigenfrequencies $f_0$ at pressures below $10^{-4}$\,mbar and at room temperature. To this end, a shear piezo is glued to the nanowire chip for piezo-actuation~\cite{bib:Favero2009a} as indicated in the inset of Fig.~2.
Nanowire vibration with amplitudes in the range of a few nanometers
is probed via direct optical detection\,\cite{bib:Sanii2010} by detecting the reflected light of a laser ($\lambda = 405$\,nm) vertically focused on the device\,\cite{bib:Paulitschke}, resulting in a displacement sensitivity of $6$\,pm$/ \sqrt{\mathrm{Hz}}$. The obtained resonance spectrum is fitted with a Lorentzian to extract both eigenfrequency and quality factor. The eigenfrequencies of the investigated nanowires cover the frequency range between $200$\,kHz and $42$\,MHz. The observed quality factors lie between $1,000$ and $4,000$, while generally higher quality factors are observed for increasing foot radii. A typical dataset featuring an eigenfrequency of $5.45$\,MHz and a quality factor of $1,500$ is depicted in Fig.~2.

\begin{figure}[htb]
\begin{center}
\includegraphics[width=8.25cm]{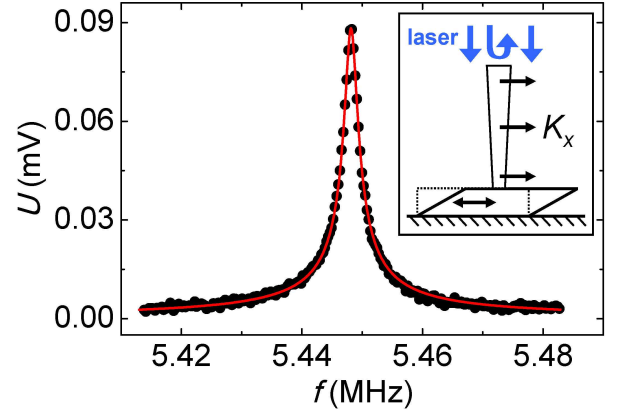}
\caption{Typical resonance curve of a nanomechanical nanowire resonator with $h=9\,\mu$m, $r=564$\,nm and $R=695$\,nm in the linear response regime. From the Lorentzian fit included as solid line a fundamental nanowire eigenfrequency $f_0 = 5.45$\,MHz as well as a quality factor of $1,500$ are determined. The inset displays a schematic of the experimental setup, including actuation via a shear piezo and optical displacement detection.}
\end{center}
\label{resonance}
\end{figure}

An important consequence of the gradually increasing cross section of the inverted conical nanowire is that the simple Euler-Bernoulli formula from eq.~(\ref{cylinder}) relating eigenfrequency, geometry and material properties for the cylindrical beam does not hold anymore. To correctly interpret the measured resonance frequencies, the appropriate functional relationship for an inverted conical nanowire has to be obtained. This is accomplished by solving the general Euler-Bernoulli differential equation~\cite{bib:Weaver1990}
\begin{equation}
\label{diff}
E\frac{d^2}{dz^2}\left(I(z)\frac{d^2X(z,t)}{dz^2}\right)= -\rho S(z)\frac{d^2X(z,t)}{dt^2},
\end{equation}
with $z$ being the axis along the nanowire length, $X(z,t)$ the eigenfunction of the bent nanowire, $\rho$ the mass density, $I(z)$ the area moment of inertial and $S(z)$ the cross sectional area of the non-prismatic nanowire at height $z$. For the inverted conical nanowire, $I(z) = \pi/4\cdot(r+\Delta(z))^4$ and $S(z) = \pi (r+\Delta(z))^2$ with $\Delta(z)$ = $z \cdot \tan \varphi$. Under these assumptions, a general solution of eq.~(\ref{diff}) can be obtained using Kirchhoff's approach~\cite{bib:Kirchhoff1879} which is constructed from Bessel functions~\cite{bib:Conway1946,bib:Conway1965}. However, this method does not enable an analytic solution of the corresponding boundary value problem. For the boundary conditions of the inverted conical nanowire,
\begin{align}
\label{bound}
X(0) = 0,\, dX/dz|_{z=0} = 0,\, \nonumber\\
d^2X/dz^2|_{z=h} = 0,\,
d^3X/dz^3|_{z=h} = 0,
\end{align}
only tabulated numerical values for selected nanowire dimensions are found in literature~\cite{bib:Conway1965}.

However, for the case of the piezo-actuated, inverted conical nanowire an analytic solution can be derived which is detailed in the supplemental material~\cite{bib:SI}.
The result can be summarized as
\begin{equation}
\label{omegasimple}
\omega_0^2 = E/\rho \cdot G^2,
\end{equation}
where the function $G$ is a real number describing the full geometry dependence of the nanowire (see eq.~(S4) of the supplemental material.\,\cite{bib:SI}).
We would like to note that this solution is valid for all aspect ratios $r/h < 0.1$ and for positive as well as negative angles $\varphi$. In the limit $\varphi \rightarrow 0$ the result approximates the well-known $G = 1.758 \cdot r/h^2$ of the singly-clamped cylinder (c.f. eq.~(\ref{cylinder})). The numerical solutions for four pre-selected aspect ratios given in the literature~\cite{bib:Conway1946,bib:Conway1965} are in almost perfect agreement with the respective values of eq.~(\ref{omegasimple}).

Equation~(\ref{omegasimple}) allows to determine the nanowires' Young's modulus $E$ from the experimentally determined eigenfrequencies $\omega_0 = 2\pi f_0$ and geometry factors $G$. Figure~3
displays the optically measured eigenfrequencies of the nanowires' fundamental flexural mode as a function of the geometry factor $G$ determined from the nanowire dimensions $r$, $R$ and $h$ obtained via SEM analysis. The data displays a linear slope across the entire range of $G$, covering a frequency range from $200$\,kHz to $42$\,MHz. This implies that the complete set of nanowires down to the most flexible ones (see inset for a magnified view of datapoints including error bars) is characterized by the same, global value of Young's modulus $E$. Using a constant mass density $\rho = 5307$\,kg$/$m$^3$ for GaAs~\cite{bib:Bateman1959}, the fitted line through origin included in Fig.~3
as solid line yields $E = (45 \pm 3)$\,GPa.

In the following, the observed global value of $E$ is compared to the existing literature. For the presented nanowires etched into ($100$)-GaAs, the flexural vibrations are governed by Young's modulus along the ($100$)-direction~\cite{bib:Hopcroft2010} which has a bulk value of $85.6$\,GPa~\cite{bib:Burenkov1973}. Lacking a reference value for nanostructures along this crystal direction we compare our results to GaAs nanowires epitaxially grown along the ($111$)-direction, for which a broad spectrum of literature can be found. Experimentally obtained values range from $37-183$\,GPa for nanowires with diameters between $55$\,nm and $160$\,nm~\cite{bib:Wang2011b,bib:Kallesoe2012,bib:Alekseev2012},
yielding a somewhat unspecific picture compared to a bulk value of Young's modulus for (111)-GaAs of $141.2$\,GPa~\cite{bib:Burenkov1973}.
\begin{figure}[htb]
\begin{center}
		\includegraphics[width=8.25cm]{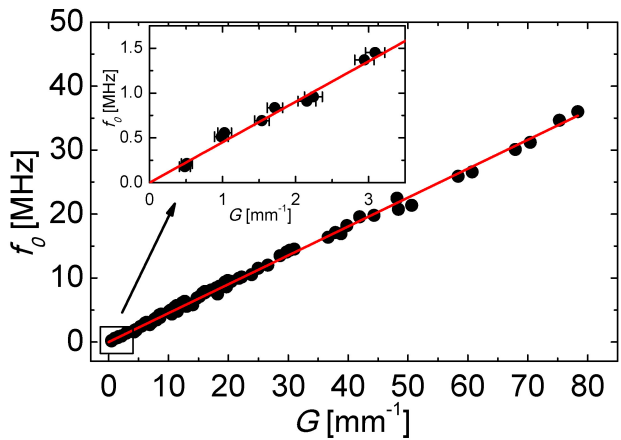}
\caption{Experimental nanowire resonance frequencies (black circles) plotted as a function of the geometry factor $G$ and fitted line through origin (solid line), yielding a constant Young's modulus $E = (45 \pm 3)$\,GPa. The inset shows an enlarged view of the leftmost part of the plot indicated by a black rectangle, clearly demonstrating that the ten most flexible nanowires are equally well-described by the global value of $E$ as their more rigid counterparts exhibiting a larger geometry factor $G$. Error bars on the $G$-axis include the uncertainty in determining $r$, $R$ and $h$ by SEM analysis.}
\end{center}
\label{emodul}
\end{figure}

A possible explanation for this vast spread could be a dependence of Young's modulus on the geometric dimensions of the underlying nanowires, which is actively discussed in part of the research literature~\cite{bib:Li2003,bib:Chen2006,bib:Gavan2009a,bib:Lexholm2009,bib:Nam2006,bib:Ngo2006,bib:Zhang2008a,bib:Loeffler2011}.
In order to elucidate possible size effects in our data, Young's modulus is additionally determined for each individual nanowire, again using eq.~(\ref{omegasimple}), now applied to the respective datapoints in Fig.~3
rather than fitting their overall slope. The resulting values denoted $\tilde E$ are plotted as a function of $G$ in the left part of Fig.~4 as black circles.
A histogram of the observed values of $\tilde E$ shown in the right part of Fig.~4 yields a normal distribution with expected value $\overline{\tilde E} = (44 \pm 6)$\,GPa (dotted line) which agrees well with the global value $E = (45 \pm 3)$\,GPa.
However, an increasingly broad distribution of $\tilde E$ values is observed in the left part of Fig.~4 for decreasing geometry factor $G$. In order to rule out a possible influence of size effects, the nanowires are segmented into five subsets containing $15$ datapoints each, which are indicated by alternating background color outlining the increasing relative error in determining the nanowires' dimensions for decreasing $G$.
For each of these subsets the above statistical analysis is repeated, yielding both average $\overline{\tilde E}_i$ ($i=1,...,5$) and standard deviation, indicated by filled squares including error bars. Clearly, even though the errors strongly increase for the subsets featuring smaller $G$, all $\overline{\tilde E}_i$ agree with $\overline{\tilde E}$ within their error limits. Furthermore the observed values of $\overline{\tilde E}_i$ are statistically distributed around $\overline{\tilde E}$ and there is no systematic increase or decrease for smaller $G$. Thus we conclude that the determined Young's modulus is not dependent on the geometric dimensions of the nanowires and remains constant even for the smallest nanowire foots of $r=36$\,nm. The increasingly broad scatter of datapoints for decreasing nanowire size (and thus $G$) is thus fully attributed to the increasing relative error in determining $r$, $R$ and $h$ by SEM inspection.
\begin{figure}[htb]
\begin{center}
		\includegraphics[width=8.25cm]{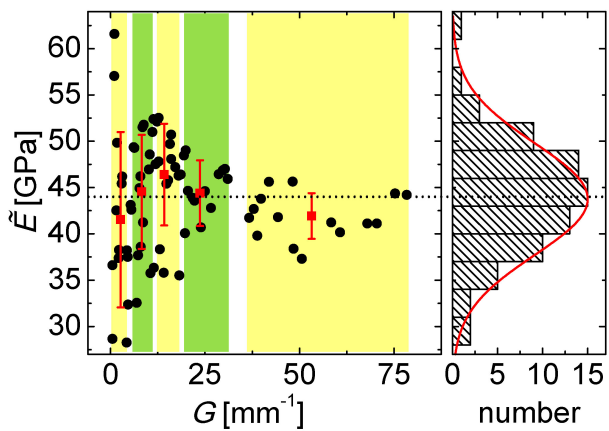}
\caption{Left: Individual values for Young's modulus $\tilde E$ (black circles) extracted for each nanowire, i.e. from each datapoint in Fig.~3, as a function of $G$. The dotted line represents the expected value $\overline{\tilde E} = 44$\,GPa. Subsets of 15 nanowires of similar $G$ are idicated by alternating background color. For each subset, the resulting values of $\overline{\tilde E}_i$ ($i=1,...,5$) as well as their standard deviation are added as filled squares including error bars. Right: Histogram and normal distribution of the observed values of $\tilde E$, yielding an expected value of $\overline{\tilde E} = (44 \pm 6)$\,GPa.}
\end{center}
\label{etilde}
\end{figure}

In conclusion, we investigate singly clamped inverted conical nanowires with lengths between $2$ to $9\,\mu$m, foot radii between $36$\,nm and $935$\,nm and a taper angle of about $1.5^\circ$ that have been etched into a ($100$)-oriented GaAs wafer by inductive coupled plasma reactive ion etching. The nanomechanical eigenfrequencies of the fundamental flexural nanowire vibration is experimentally characterized, spanning two orders of magnitude in frequency from $200$\,kHz to $42$\,MHz. This comprehensive set of inverted conical nanowire resonators allows for a detailed analysis of this previously ill-explored type of nanomechanical system. Along with an analytic eigenfrequency solution derived from the Euler-Bernoulli boundary value problem for the inverted conical nanowire described by equations~(\ref{diff}) and (\ref{bound}), a single, global value for Young's modulus of $E = (45\pm 3)$\,GPa is determined. This value applies to all investigated nanowire geometries, and does not exhibit size effects.

The presented work sets the foundations for further investigations of inverted conical nanowires as nanomechanical systems. The combination of an optically detectable nanowire head with an extremely narrow nanowire foot makes inverted conical nanowires a promising candidate for ultra-sensitive devices, capable to probe minute forces exerted on a nanowire via its head deflection~\cite{bib:Melli2010}. Precise process control and top down nanofabrication techniques allow to tailor the nanowires' spring constant for the desired application~\cite{bib:Waggoner2007,bib:Arlett2011} and simultaneously facilitates the realization of large, custom-coupled nanowire arrays~\cite{bib:Spletzer2008}. Besides future applications as force sensors, such ultraflexible nanowire arrays are ideal model systems to explore the nonlinear dynamics and coupling phenomena in nanomechanical arrays~\cite{bib:Buks2002,bib:Lifshitz2012}, for which synchronization~\cite{bib:Perlikowski2008a} and Q-boosting~\cite{bib:Lin2007a} have been predicted. Metallized nanowire heads will allow to integrate plasmonic functionality~\cite{bib:Bora2010} or a transport degree of freedom~\cite{bib:Wiersig2008}. Finally, the integration of photonic elements such as Bragg reflectors or  quantum dots can be envisioned~\cite{bib:Lauhon2004,bib:Schneider2012a}, which may lead to additional functionality incorporating optomechanical arrays~\cite{bib:Heinrich2011,bib:Xuereb2012f}.

\begin{acknowledgments}
Financial support from the Volkswagen Foundation, the German Excellence Initiative via the Nanosystems Initiative Munich (NIM), LMUinnovativ as well as LMUexcellent, and the Center for NanoScience (CeNS) is gratefully acknowledged. The authors appreciate ongoing support and stimulating discussion with J.~P.~Kotthaus.
\end{acknowledgments}

\onecolumngrid
\clearpage
\newcommand{\fig}[1]{Fig.\,\ref{#1}}
\newcommand{\eq}[1]{equation~(\ref{#1})}

\renewcommand{\figurename}{Supplementary Figure}
\renewcommand{\thefigure}{S\arabic{figure}}
 \renewcommand{\theequation}{S\arabic{equation}}
 \setcounter{figure}{0}
\setcounter{equation}{0}
\renewcommand{\fig}[1]{Supplementary Figure~\ref{#1}}
\renewcommand{\eq}[1]{equation~(\ref{#1})}

\section*{Supplementary material:\\ Euler-Bernoulli theory for the inverted conical nanowire}

\noindent The following discussion is dedicated to the solution of the Euler-Bernoulli problem of an inverted conical mechanical resonator, which is singly clamped at its narrow end. Its equation of motion, the
general Euler-Bernoulli differential equation~\,[1]
\begin{equation}
\label{diff}
E\frac{d^2}{dz^2}\left(I(z)\frac{d^2X(z,t)}{dz^2}\right)= -\rho S(z)\frac{d^2X(z,t)}{dt^2},
\end{equation}
is included in the main text as eq.\,(2). Here, $z$ denotes the axis along the nanowire length, $X(z,t)$ the eigenfunction of the bent nanowire, $\rho$ the mass density, $I(z)$ the area moment of inertial and $S(z)$ the cross sectional area of the non-prismatic nanowire at height $z$. For the inverted conical nanowire, both the area moment of inertia $I(z) = \pi/4\cdot(r+\Delta(z))^4$ and the cross sectional area $S(z) = \pi (r+\Delta(z))^2$ depend on a linear radius increment $\Delta(z)$ = $z \cdot \tan \varphi$ using the designation assigned in the inset of Fig.\,1 of the manuscript.

\subsection*{Solving the boundary value problem for the case of distributed uniform loading}
\noindent The boundary conditions of the inverted conical nanowire
\begin{equation}
\label{bound}
X(0) = 0,\, dX/dz|_{z=0} = 0,\, d^2X/dz^2|_{z=h} = 0,\, d^3X/dz^3|_{z=h} = 0,
\end{equation}
are summarized in eq.\,(3) of the manuscript. In the following, an analytic solution of equations~(\ref{diff}) and (\ref{bound}) will be derived for the piezo-actuated inverted conical nanowire.
The right hand side of eq.~(\ref{diff}) represents a force per unit length acting on the cross section of the wire $S(z)$ at height $z$. Assuming this force is mediated by a shear piezo (see inset of Fig.~2 of manuscript),
its position-dependent part can be considered as $K_x = \mathrm{const}$, eliminating the $z$-dependence of the actuating force. This assumption of a distributed uniform load is justified as long as the condition $f_0 \cdot \tau \ll 1$ is fulfilled for the acoustic travel time $\tau = h/v_T$ along the nanowire length $h$, with $v_T = 3,345$\,m/s being the velocity of sound of the transverse wave in ($100$)-direction in GaAs~\,[2]. In other words, $K_x$ is constant as long as the retardation time $\tau$ until the free end of the nanowire experiences the acceleration of the driven substrate is negligibly small compared to the vibration period of the resonator.
We would like to note that this condition is generally fulfilled for nanoresonators with $r/h<0.1$ regardless of the underlying material's $v_T$.

For the special case of a distributed uniform load, i.e. a constant $K_x$, the boundary value problem consisting of equations~(\ref{diff}) and (\ref{bound}) is solved analytically
for a singly clamped inverted conical nanowire with $\varphi \neq 0$ using Wolfram Research Mathematica, yielding the eigenfunction for the fundamental flexural mode
\begin{eqnarray}
\label{dx}
X_0(z) &=& \frac{K_x}{3 b^4 E \pi  r^3 (r+b z)^2}  \left\{F_1 + F_2\right\} \\
F_1 &=& b z \left [ 6 r^4+3 b r(b^2 h^2+3 r^2)z +2 b^2(b^2 h^2-b h r+r^2) z^2\right ] \nonumber \\
F_2 &=& 6 r^3 (r+b z)^2 \left [\log(r)-\log(r+bz)\right ]\nonumber ,
\end{eqnarray}

with $b \mathrel{\mathop:\!\!=} \tan \varphi$. Note that in the limit $\varphi \rightarrow 0$ the solution converges to the well-known result for the cylindrical beam~\,[3] $X_0(z) = K_x/(6E\pi r^4) \cdot \left [x^2 (6h^2-4hx+x^2)\right]$.

\subsection*{Determining the eigenfrequency}
\noindent The eigenfrequency $f_0$ of the inverted conical nanowire is determined by equating the maximum potential energy $U_{max}=\frac{1}{2}E\int^{h}_{0}I(z)\left(\frac{\partial^2X_0(z)}{\partial z^2}\right)^2 dz$ and the maximum kinetic energy $T_{max}=\frac{\omega_0^2}{2}\int^{h}_{0} \rho S(z) X_0(z)^2 dz$ of the conservative system. For a harmonic oscillator with $X_0(z,t)=X_0(z) \sin(\omega_0 t + \phi)$ this results in the following analytic expression for the angular eigenfrequency $\omega_0= 2 \pi f_0$ of the nanowire
\begin{eqnarray}
\label{omega}
\omega_0^2 &=&\frac{E}{\rho}\cdot \underbrace{
\frac{90b^4r^3 [ bh F_3 + 12 r^3 F_4 ]}{bh [F_5 + F_6] + 60r^3 (bh + r)^3 F_7  (\log(r) - \log(bh + r))}}_{\text{\large $:=G^2$}}
\\[1.1ex]
F_3 &=& b^3h^3 - 2b^2h^2r + 6bhr^2 + 12r^3 \nonumber \\
F_4 &=& (bh + r)(\log(r) -\log(bh + r)) \nonumber \\
F_5 &=& 24b^8h^8 - 18b^7h^7r + 72b^6h^6r^2 + 207b^5h^5r^3 - 21b^4h^4r^4 \nonumber \\
F_6 &=& 375b^3h^3r^5 +1490b^2h^2r^6+ 1410bhr^7 + 420r^8 \nonumber \\
F_7 &=& 3b^3h^3 - 6b^2h^2r + 12bhr^2+ 7r^3 + 6r^3(\log(r)- \log(bh + r))\nonumber .
\end{eqnarray}

\begin{figure}[htb]
\begin{center}
		\includegraphics[width=11cm]{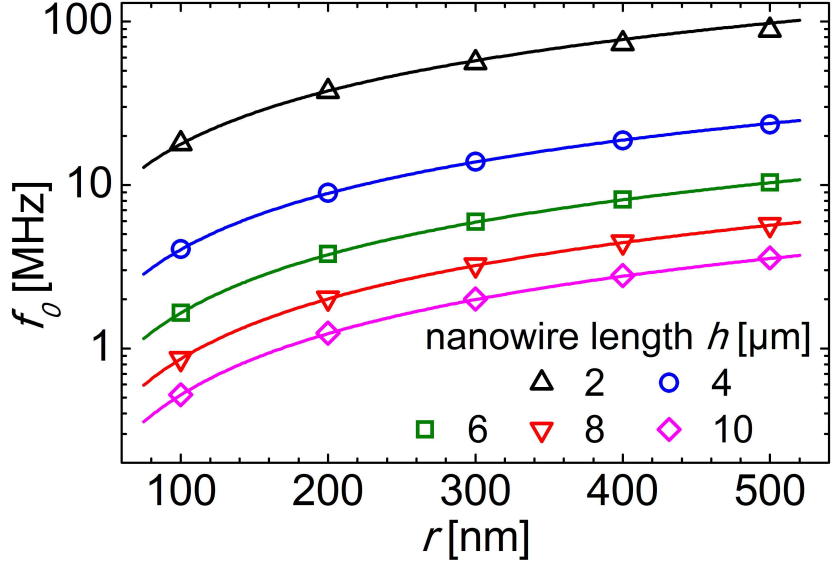}
\caption {Comparison of analytically and numerically determined eigenfrequencies. Solid lines depict analytically determined eigenfrequencies using eq.~(\ref{omega}) as a function of the nanowire foot radius $r$ for five nanowire lengths $h$. Hollow symbols display eigenfrequencies simulated using COMSOL Multiphysics for different values for $r$ and $h$, assuming $E = 45$\,GPa and $\rho = 5307$\,kg$/$m$^3$. For both analytical and numerical analysis, a fixed taper angle of $\varphi = 1^\circ$ has been employed. There are no free parameters.}
\end{center}
\label{fem}
\end{figure}
Figure~S1
displays the resulting eigenfrequencies of the fundamental mode as a function of $r$ for a set of nanowire lengths $h$ (solid lines). The analytically derived solutions are complemented by numerical eigenfrequency analysis for the experimentally explored nanowire geometries using COMSOL Multiphysics (hollow symbols).
Excellent agreement between the analytical solution from eq.~(\ref{omega}) and the numerical simulations is found for the full range of geometric parameters explored, corresponding to nanowire eigenfrequencies spanning more than two orders of magnitude. The maximum relative deviation remains below $2\,\%$ for nanowire aspect ratios $r/h < 0.1$ and moderate taper angles $\varphi$ between $0^\circ$ to $3^\circ$ (dependence on taper angle not shown).
Only for aspect ratios $r/h > 0.1$, the numerical eigenfrequency results start differing from eq.~(\ref{omega}), as apparent for the rightmost upright hollow triangle corresponding to $r=500$\,nm and $h=2\,\mu$m which exhibits a $10\%$ deviation from the solid line. This is a consequence of the increasing influence of shearing deformations~\,[1], which go beyond the standard Euler-Bernoulli beam theory approach and are not included in the above derivation.

\subsection*{Supplementary references}
\setlength{\parindent}{0pt}

\begin{tabular}{ll}
{[1]} & W. Weaver, S. P. Timoshenko, and D. H. Young, {\it Vibration problems in engineering}, 5th ed.
(John Wiley \& \\
 & Sons, 1990).\\
{[2]} & T. B. Bateman, H. J. McSkimin, and J. M. Whelan, Journal of Applied Physics 30, 544 (1959).
\\
{[3]} & A. N. Cleland, {\it Foundations of Nanomechanics} (Springer Verlag Berlin Heidelberg New York,
2003).\\
\end{tabular}

\end{document}